# Parameter estimation for WMTI-Watson model of white matter using encoder-decoder recurrent neural network

Yujian Diao[1,2,*] and Ileana Ozana Jelescu[3]

[1]Laboratory of Functional and Metabolic Imaging, Ecole Polytechnique Fédérale de Lausanne, Lausanne, Switzerland
[2]CIBM Center for Biomedical Imaging, Lausanne, Switzerland
[3]Department of Radiology, Lausanne University Hospital, Lausanne, Switzerland

*Corresponding author: Yujian Diao (yujian.diao@epfl.ch)

**Abstract**

Biophysical modelling of the diffusion MRI signal provides estimates of specific microstructural tissue properties. Model parameters estimates can be obtained by fitting the model to the measured signal. Although nonlinear optimization such as non-linear least squares (NLLS) is the most widespread method for model estimation, it suffers from local minima, high computational cost and uncertain accuracy. Deep Learning approaches are steadily replacing NL fitting, but come with the limitation that the model needs to be retrained for each acquisition protocol and noise level. The White Matter Tract Integrity (WMTI)-Watson model was proposed as an implementation of the Standard Model of diffusion in white matter that estimates model parameters from the diffusion and kurtosis tensors (DKI), thereby overcoming fitting the model signal equation. Here we proposed a deep learning approach based on the encoder-decoder recurrent neural network (RNN) to increase the robustness and accelerate the parameter estimation of WMTI-Watson. We use an embedding approach to render the model insensitive to potential differences in distributions between training data and experimental data. This RNN-based solver thus has the advantage of being highly efficient in computation and more readily translatable to other datasets, irrespective of acquisition protocol and underlying parameter distributions as long as diffusion and kurtosis tensors (or their typical derived scalars) were pre-computed from the data. In this study, we evaluated the performance of NLLS, the RNN-based method and a baseline DL architecture based on multilayer perceptron (MLP) on synthetic and in vivo datasets of rat and human brain. We showed that the proposed RNN-based fitting approach had the advantage of highly reduced computation time over NLLS (from hours to seconds), with similar accuracy and precision but improved robustness, and superior translatability to new datasets over MLP, irrespective of acquisition protocol or species being rat or human.

**1. Introduction**

Diffusion magnetic resonance imaging (dMRI) which encodes information about brain white matter (WM) microstructure in diffusion-weighted signal has emerged in recent years as a highly promising technique to provide specific information about microstructure features. One major approach of decoding tissue information and retrieving specific microstructure features from diffusion signal is biophysical modelling. Assuming a simplified geometry of the underlying tissue, biophysical models express the diffusion signal in the chosen environment in analytical equations and estimate relevant parameters of the microstructure by fitting to the dMRI data (Jelescu et al., 2020; Jelescu and Budde, 2017; Novikov et al., 2019). As to model estimation, most models employ nonlinear (NL) fitting such as nonlinear least squares (NLLS) that is strongly affected by noise and model degeneracy which means the existence of multiple sets of solutions due to local minima or the flat fitting landscape (Harms et al., 2017; Jelescu et al., 2016; Novikov et al., 2018). To mitigate degeneracy, current models usually implement additional constraints, noise models and initialization strategies which result in decreased generalization and incompatibility between different models. Furthermore, NL fitting is notoriously expensive and slow in computation preventing it from being applied in large scale datasets.

Recently, deep Learning (DL) based approaches have been proposed to replace NL fitting and achieved promising outcomes (Barbieri et al., 2020; de Almeida Martins et al., 2021; Park et al., 2021; Ye, 2017; Ye et al., 2019, 2020). However, when based on fully connected feedforward networks or convolutional neural networks (CNN), DL approaches are mostly trained directly on the diffusion signal and sometimes take the training target as parameter estimates from conventional fitting methods (e.g. NLLS) instead of the ground truth (Park et al., 2021; Ye et al., 2020). Such implementations limit the generalization ability of the DL models to dMRI signals with different noise levels or acquired with different protocols (b-values, directions…) since the models are highly dependent on the underlying parameter distributions in the training set, which may vary across species, disease or brain regions.

Other than estimating parameters directly from the dMRI signal, the White Matter Tract Integrity (WMTI) model (Fieremans et al., 2011) and its extension to a Watson distribution of axon orientations (Jespersen et al., 2018) were proposed as an implementation of the Standard Model of diffusion in WM that estimated model parameters from the diffusion and kurtosis tensors (DKI), thereby overcoming fitting the model signal equation. This approach has two main advantages. First, DKI is a clinically-feasible technique where the tensors can be computed from the diffusion signal using linear least squares (Jensen et al., 2005; Veraart et al., 2013), and is now part of routine dMRI processing pipelines such as MRtrix (Tournier et al., 2019) and DESIGNER (Ades-Aron et al., 2018). Second, building on DKI empirical scalar metrics (mean/axial/radial diffusivities and kurtoses), the WMTI-Watson model computes microstructure-specific parameters of WM, such as axonal water fraction, orientation dispersion and compartment-specific diffusivities without additional constraints or fixed parameters. In this study, we propose a novel DL approach based on encoder-decoder recurrent neural network (RNN) which has been very successful in natural language translation (Cho et al., 2014; Sutskever et al., 2014) to predict WMTI-Watson model parameters from DKI with robustness and acceleration as compared to NLLS. The RNN estimator was trained on synthetic data with ground truth which was sampled following realistic distributions drawn from in vivo rat brain data. Moreover, as DL models assume similarity of test and training distributions, we employ a distribution matching technique which embeds new test data in the training data to enable straightforward model applicability to data with different distributions. The strength of the proposed RNN estimator is to enable WMTI-Watson model parameter estimation on any dataset, provided DKI was pre-computed from the dMRI signal. It should be noted however that the DKI tensor estimations themselves depend on the acquisition protocol (e.g. the two non-zero b-values, number of directions per shell etc.), although substantial harmonization efforts are underway.

In this study, we evaluated the proposed RNN model on synthetic data, in vivo rat data as well as human data. We also compared its performance to NLLS as well as to another DL solver based on a fully-connected feedforward network (Multi-layer perceptron – MLP).

## 2. Methods

### 2.1. Biophysical model

WMTI-Watson is a declination of the Standard Model. It is comprised of two non-exchanging Gaussian compartments (intra- and extra-axonal) where axons are modelled by zero radius sticks embedded in a Gaussian anisotropic extra-axonal space (Fieremans et al., 2011; Jespersen et al., 2018). The intra-axonal space is characterized by a volume fraction of water molecules $f$ and intra-axonal diffusivity $D_a$ while the extra-axonal space is described by extra-axonal axial and radial diffusivities, $D_{e,\parallel}$ and $D_{e,\perp}$ (Figure 1). The fiber orientation distribution function (fODF) of the axons is chosen to be a Watson distribution which can be narrowed down to a single metric, the mean cosine-squared of axon orientations with respect to the main bundle orientation: $c_2 = \langle\cos(\Psi)^2\rangle$, with $c_2 = \frac{1}{3}$ corresponding to isotropically distributed axons and $c_2 = 1$ to perfectly parallel axons for an axially-symmetric fiber tract. Generally, the diffusion signal S is given by (Jespersen et al., 2018; Novikov et al., 2018; Reisert et al., 2017):

$$S(b, \hat{n}) = \int \mathcal{P}(\hat{u})\left(fe^{-bD_a(\hat{u}.\hat{n})^2} + (1-f)e^{-bD_{e,\parallel}(\hat{u}.\hat{n})^2 - bD_{e,\perp}(1-(\hat{u}.\hat{n})^2)}\right)d\hat{u}$$

Basically, it is a combination of signal originating from the intra- and extra-axonal tissue integrated by the fODF $\mathcal{P}$ across all directions. For a fODF that is modelled by a single parameter, all five model parameters can be directly derived from diffusion and kurtosis tensors (Jespersen et al., 2018; Kunz et al., 2018). Hence model parameters **p** = $[f, D_a, D_{e,\parallel}, D_{e,\perp}, c_2]$ can be estimated exploiting the stable linear fitting of DKI tensors: mean/axial/radial diffusivities (MD/AD/RD) and mean/axial/radial kurtosis (MK/AK/RK). This model has two mathematical solutions, which can be discriminated by either $D_a < D_{e,\parallel}$, or $D_a > D_{e,\parallel}$ (Fieremans et al., 2011; Jelescu et al., 2016; Novikov et al., 2018). Here, we choose to train the models towards $D_a > D_{e,\parallel}$ according to recent evidence (Dhital et al., 2019; Howard et al., 2020; Jespersen et al., 2018; Kunz et al., 2018). In addition, DKI and WMTI model parameters are constrained by the following physical bounds: 0 < MD, AD, RD < 3; 0 < MK, AK, RK < 10; 0 < $f$ < 1; 0 < $D_a$ < 4; 0 < $D_{e,\parallel}, D_{e,\perp}$ < 3; 1/3 < $c_2$ < 1. Notably, the upper bound for $D_a$ was relaxed to 4 $\mu m^2/ms$ instead of 3 which is the free diffusion coefficient at 37 °C because the expected $D_a$ value is relatively high and noise may push it towards 3. Setting a strict upper bound at 3 would therefore bias the outcome.

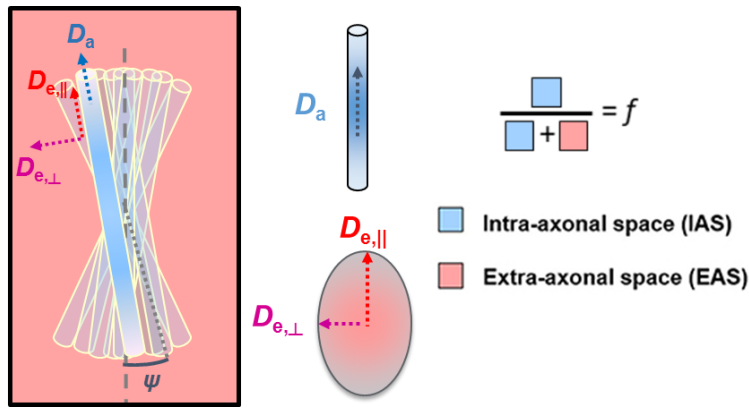

Figure 1. Schematic of the two-compartment WMTI-Watson model with relevant parameters *(Jelescu and Budde, 2017)*. $D_a$ [$\mu m^2/ms$] is intra-axonal diffusivity. $D_{e,\parallel}$ and $D_{e,\perp}$ [$\mu m^2/ms$] are local extra-axonal diffusivities. $f$ is axonal volume fraction. Orientation dispersion is reported as $c_2 = \langle cos(\Psi)^2\rangle$.

### 2.2 Neural network architecture and training

In this work, we formulate the estimation of model parameters as a sequence-to-sequence (seq2seq) prediction problem since five model parameters are to be estimated from six DKI metrics, which can each be viewed as two sequences. We note that the six DKI metrics can be reduced to five independent ones (since $3MD = AD + 2RD$), but have chosen a sequence of six to keep consistent with NLLS fitting. We therefore propose a deep neural network based on the RNN encoder-decoder architecture (Cho et al., 2014) with DKI metrics as the input sequence and the WMTI-Watson parameters as the output sequence (Figure 2). As comparison, we also designed a baseline DL model based on MLP, which is a fully-connected feedforward neural network.

As the name suggests, the RNN encoder-decoder model is composed of two RNN sub-networks - Encoder and Decoder, each consisting of a series of Long Short-Term Memory (LSTM) cells for encoding and decoding the sequence. Specifically, the encoder learns a representation of the entire input sequence that captures its characteristics by extracting key features from the given data into a vector known as hidden state or context vector. This context vector which aims to contain a good summary of input elements is passed along to the decoder as input to help the decoder make accurate predictions. The decoder receives the context vector which is the last hidden state of encoder as its initial decoder hidden state and starts decoding the output sequence one by one. At each step, the decoder produces an output element as well as a hidden state which are used to predict the element at next step. When only the context vector is passed to decoder from encoder, it might have difficulty to compress the contextual information of the entire sentence into a single fixed length vector especially when sequences are long (Bahdanau et al., 2016). Attention mechanism was therefore proposed as a solution by learning the alignment of the input sequence to the output sequence to allow the decoder to focus on the most relevant parts of the input sequence when predicting output at each step (Bahdanau et al., 2016; Luong et al., 2015). Hence, in addition to the context vector the encoder also passes a matrix composed by the hidden vectors from each of its LSTM units to the decoder. By learning a set of attention scores multiplied by the encoder hidden vectors, the decoder will be able to give selective attention to specific part of input sequence and then relate them to elements in the output sequence.

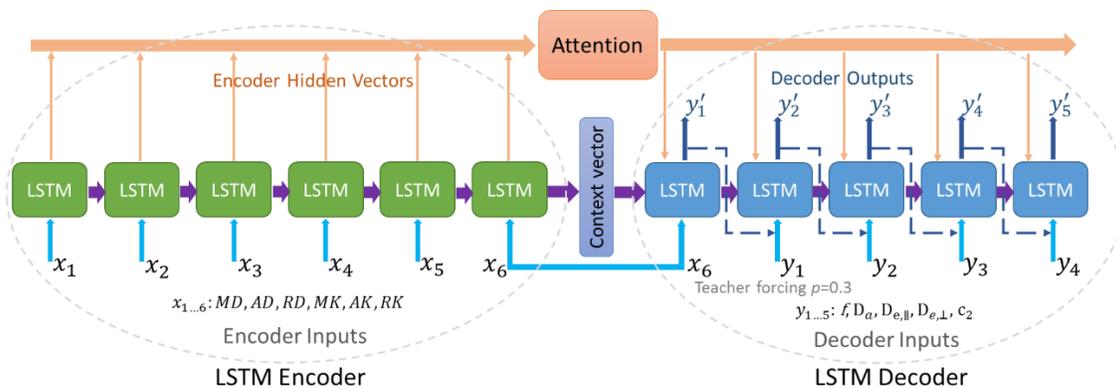

Figure 2. Architecture of the RNN encoder-decoder network. It consists of an encoder and a decoder, both based on LSTM units. The encoder encodes the 6-element input sequence (MD, AD, RD, MK, AK and RK) into context vector. Decoding is an iterative process in which the decoder takes the hidden state and element from the previous step to predict next output element one by one. For better performance, attention mechanism and teacher forcing *(Goodfellow et al., 2016)* are applied in the encoder-decoder network. The teacher forcing ratio here is 0.3 which means with 30% probability the decoder takes the ground truth $y_t$ to calculate the next element $y'_{t+1}$ in the output sequence otherwise it uses the decoder output $y'_t$ to predict the next element $y'_{t+1}$.

In the RNN network, the encoder and decoder consist of $L$ = 6 and 5 cascaded LSTM cells respectively, corresponding to the length of input sequence (MD, AD, RD, MK, AK, RK) and output sequence ($f, D_a, D_{e,\parallel}, D_{e,\perp}, c_2$). Each LSTM cell has $h$ = 96 features in the hidden state or vector. The attention mechanism is implemented by a feed-forward layer, using the decoder's input and hidden states as inputs to calculate attention weights. In addition, batch and layer normalization are applied in both encoder and decoder. Furthermore, teacher forcing technique was used in training, which means the decoder will take as input the ground-truth element instead of the output from a prior step, with a certain probability (Goodfellow et al., 2016; Williams and Zipser, 1989). The probability, referred to as a teacher forcing ratio, was chosen to be 0.3.

The MLP baseline is formed of an input layer, three hidden layers and an output layer. As a result of trial and error, hidden layers were constructed with 2048, 1024 and 256 nodes respectively and LeakyReLU (Rectified Linear Unit) was chosen as the activation function (He et al., 2015; Jelescu et al., 2022). The output layer comprised 5 nodes, each providing an estimation of the WMTI-Watson model parameters ($f, D_a, D_{e,\parallel}, D_{e,\perp}, c_2$).

For both RNN and MLP networks, supervised training was performed using a mean squared error loss (MSE)

$$MSE = \|y - y'\|^2$$

Where $y$ is the ground-truth vector and $y'$ is the corresponding network output. They were trained using mini-batches with a batch size of 256, and an Adam optimizer. For better performance, the input DKI and output WMTI-Watson parameters were both rescaled between 0 and 1 and then multiplied by 100 using a min-max normalization policy before being fed to the networks. The neural networks were implemented using Pytorch (v1.8.1) on a HP workstation hosting a NVIDIA GeForce RTX 2080 Ti graphic card with 11 GB memory as well as a 32-core Intel Xeon 2.1 GHz CPU with 2 threads per core and 126 GB memory.

**2.3. In vivo data acquisition and processing**

**Rat data**

Data from a previous longitudinal study of WM microstructure alterations in a rat model of Alzheimer's disease (Tristão Pereira et al., 2021) were reused. Details of experimental procedures can be found in the original publication. Briefly, all experiments were approved by the local Service for Veterinary Affairs. Male Wistar rats (236 ± 11 g) underwent a bilateral intracerebroventricular (icv) injection of either streptozotocin (3 mg/kg, STZ group) or buffer (CTL group). When delivered exclusively to the brain, streptozotocin induces impaired brain glucose metabolism and is used as a model of sporadic Alzheimer's disease (AD) (Grieb, 2016; Lester-Coll et al., 2006).

As a result of system upgrade, the MRI data were acquired on two rat cohorts (1/2) on a 14.1 T small animal scanner with different MRI consoles (Varian/Bruker). Animals scanned on the Varian system and Bruker system are referred as *Cohort1* (N = 17 rats) and *Cohort2* (N = 7 rats), respectively. Each cohort comprised animals from both groups: Cohort1 (CTL/STZ, N = 8/9 rats) and Cohort2 (CTL/STZ, N = 4/3 rats), which were scanned at 2, 6, 13, and 21 weeks after the surgery. The same in-house built quadrature surface transceiver was used in all experiments. The acquisition parameters were the same for the two cohorts. Diffusion-weighted data were acquired using a pulsed-gradient spin-echo (PGSE) segmented echo-planar-imaging (EPI) sequence, with following parameters: 4 b=0 and three shells b=0.8/1.3/2 ms/μm² with 12/16/30 directions; δ/Δ=4/27 ms; TE/TR=48/2500 ms, 9 1-mm slices, FOV=23 × 17 mm², matrix=128 × 64, 4 repetitions.

Diffusion data processing included MP-PCA denoising (Veraart et al., 2016), Gibbs-ringing correction (Kellner et al., 2016) and correction for EPI-related distortions using FSL's eddy (Andersson and Sotiropoulos, 2016). The diffusion and kurtosis tensors were estimated using a weighted linear least squares algorithm (Veraart et al., 2013) and typical DTI and DKI-derived metrics were computed: fractional anisotropy (FA), mean/axial/radial diffusivity and mean/axial/radial kurtosis. The biophysical WMTI-Watson model was computed in white matter regions using a voxel-by-voxel NLLS fit to extract microstructure parameters: $f, D_a, D_{e,\parallel}, D_{e,\perp}$, and $c_2$. In parallel, FA maps were registered to an FA template in the Waxholm Space using linear and non-linear registration in FSL (Jenkinson et al., 2002) and the corpus callosum (CC), cingulum (CG) and fimbria (Fi) of the hippocampus were automatically segmented.

**Human data**

In this study, data from the Human Connectome Project (HCP) was used (Van Essen et al., 2012). Diffusion MRI data of 6 subjects (Female/Male, N=3/3; Age, 22-35) was randomly selected from HCP's 1200 Subject Release. The HCP data was acquired on a 3 Tesla Connectome Scanner using a spin-echo EPI sequence with the following parameters: 6 b=0 and 2 b-shells b=1/2 ms/μm² with 90 directions; TE/TR=89.5/5520 ms, 145 1.25-mm slices, FOV= 181 × 218 mm², matrix=145 × 174. Diffusion images were preprocessed using HCP's diffusion processing pipeline (Glasser et al., 2013) which included intensity normalization across runs, TOPUP for EPI distortion correction, corrections for eddy currents, motion and gradient nonlinearity, and registration to structural space.

**2.4. Training data generation**

The general idea of generating synthetic training data is to first sample the ground-truth WMTI-Watson model parameters ($f, D_a, D_{e,\parallel}, D_{e,\perp}, c_2$) from given distributions and calculate back the DKI metrics (MD, AD, RD, MK, AK and RK) associated with those model parameters using linear equations (Jespersen et al., 2018). Neural networks are prone to overfitting which means the model learns to fit the training data perfectly, resulting in reduced generalizability to new data. Hence, we chose to generate the synthetic data by sampling parameters from distributions extracted from in vivo rat brain data estimated using a NLLS fit, instead of a random sampling of uniform distributions within the parameter bounds. As a result, the training data have a higher chance to share a similar distribution with unseen test data and the trained model will be thus more likely to keep its generalization on other experimental data.

Here, DKI and WMTI model parameter maps from cohort 2 (rat data on Bruker console) were chosen as distributions to generate synthetic data. For the training dataset, we extracted two groups of parameter distributions from voxels in three WM regions of interest (ROIs, corpus callosum (CC), cingulum (CG), and fimbria (Fi)). The first group of distributions (Figure 3A) was drawn from voxels satisfying the boundary conditions: 0 < [MD, AD, RD] < 3; 0 < [MK, AK, RK] < 10; 0.001 < $f$ < 1; 0.001< $D_a$ < 3.99; 0 < [$D_{e,\parallel}$, $D_{e,\perp}$] < 3; 0.33 < $c_2$ < 1. In the second group, the distribution was drawn from voxels following the inequality $D_a > D_{e,\parallel}$ in addition to the previous boundary conditions (Figure 3B). Afterwards, 25% and 75% of the ground-truth WMTI parameters were sampled from the two groups of distribution, respectively, and then the corresponding DKI metrics were calculated. By doing so, we make the trained model favor the $D_a > D_{e,\parallel}$ outcome but still allow it to experience other possible solutions (e.g. in case the inequality does not hold in pathology). The final distribution of DKI and WMTI parameters in the training dataset can be found in Figure 3C. Here, we only chose WM voxels because the WMTI-Watson model is designed for white matter and we thus wanted to optimize the DL estimator for this tissue.

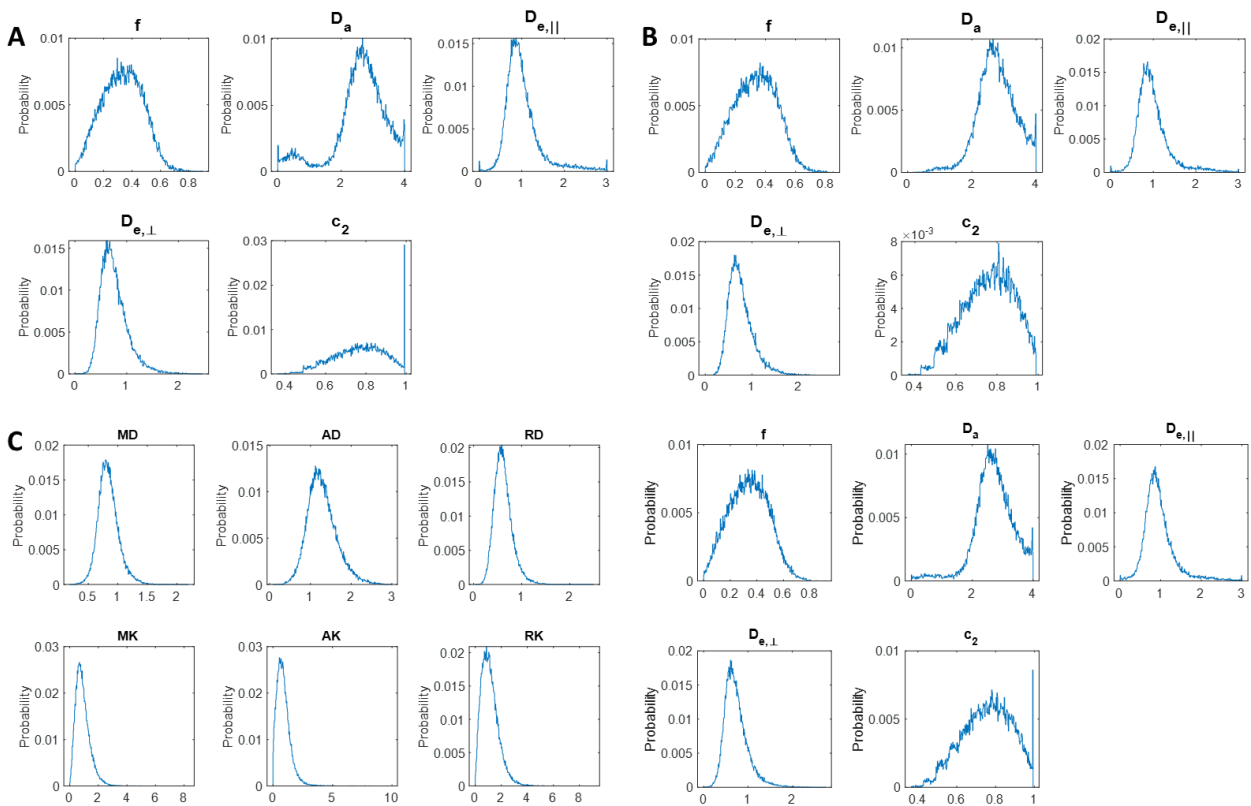

Figure 3. Distributions in WMTI-Watson parameters estimated by NLLS in rat cohort2 that were used to generate the ground-truth targets in the training data. **A.** Distributions from WM voxels which satisfies the following boundary conditions for DKI and WMTI parameters: $0 < [MD, AD, RD] < 3$; $0 < [MK, AK, RK] < 10$; $0.001 < f < 1$; $0.001 < D_a < 3.99$; $0 < [D_{e,\|}, D_{e,\perp}] < 3$; $0.33 < c_2 < 1$. 25% of targets in training data was sampled from this set of distributions. **B.** Distributions in the voxels that further satisfied $D_a > D_{e,\|}$ in addition to the boundary condition. 75% of targets was sampled from those distributions. **C.** Distributions of DKI and WMTI parameters in the synthetic training dataset. DKI parameters were computed from the ground-truth WMTI parameters.

In the end, the total synthetic dataset consisted of 4 million sequence pairs with DKI as input and WMTI parameters as output. Finally, we split this synthetic dataset into three subsets: training (80%), validation (10%) and test (10%). The validation set was used for evaluation during the training process.

### 2.5. Training and test distributions matching

DL assumes that training and test datasets are drawn from the same probability distribution and if this condition is not met, its performance will be substantially degraded. In our case, assuming a new dataset for which WMTI model parameters should be estimated without ever resorting to NLLS, only the DKI maps are available upfront. Therefore, it is not possible to filter the data based on WMTI parameter pertinence, to ensure a similar distribution to that of the training data. In fact, distributions of the NLLS-fitted in vivo data without filtering out unphysical values are quite different from the (filtered) training data distributions (Figure 4). To circumvent this problem, we embedded the new test data randomly within the training data with an embedding ratio defined as $\gamma = \frac{\text{training size}}{\text{test size}}$ (e.g. $\gamma = 10$) such that the test data and training data preserve consistent distributions. The positions at which test data are embedded are also recorded, and after the estimation the original test data will be de-embedded from those positions. The drawback of data embedding is that the ultimate test data becomes $\gamma$ times larger, which may slow down the estimation. However, in practice, this is not a problem since the DL-based estimation is extremely fast (in seconds for 1 million estimations).

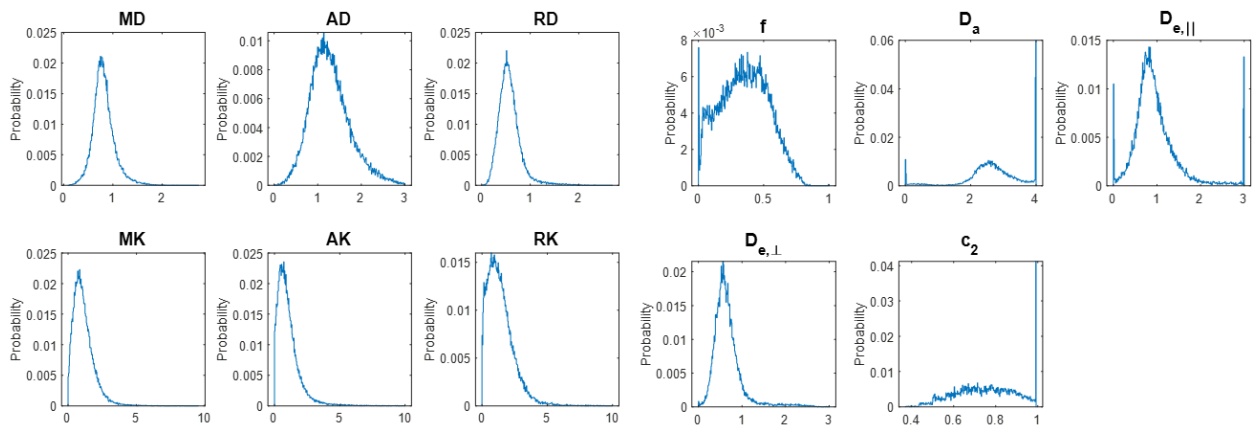

Figure 4. DKI and WMTI distributions in rat in vivo data, fitted using NLLS and only filtered by the DKI boundary conditions: *0 < [MD, AD, RD] < 3; 0 < [MK, AK, RK] < 10*. These distributions are different from the ones in the training data (Figure 3C).

## 2.6. Evaluation

**On synthetic data vs ground truth**

First, we used the unseen synthetic test data (*Testset1*, $N = 4\times10^5$), which has the same distribution as the training data, to evaluate the three fitting methods: NLLS, RNN and MLP. The estimation time of each method was also recorded. The NLLS fitting was performed using Matlab's built-in function *lsqnonlin* (more details can be found in *Supplementary Table 1*) with 32 workers in the parallel pool. RNN and MLP estimations were implemented by Pytorch on a graphic card (NVIDIA GeForce RTX 2080 Ti). We define the error between estimations and ground truth (GT) for each method as $err = \frac{Estimation - GT}{GT} \times 100\,\%$ and $err \leq 5\,\%$ is referred to as *Estimation Accuracy*. The estimation accuracy of the three methods was compared.

Secondly, to simulate concrete experimental conditions, we generated a second synthetic test dataset (*Testset2*, $N = 2\times10^5$) following the distribution (Figure 4) from data filtered only by DKI boundary conditions. Again, WMTI parameters were estimated by the three fitting methods and compared to the ground truth. Since distributions in *Testset*2 are quite different from those in training data, distribution matching by embedding ($\gamma = 10$) was employed for RNN and MLP. The error was calculated for all *Testset*2 samples, as well after filtering by ($0.01 < f < 1$; $0.01 < D_a < 3.9$; $0.01 < [D_{e,\|}, D_{e,\perp}] < 3$; $0.33 < c_2 < 1$; $D_a > D_{e,\|}$) based on estimation and ground truth values, respectively.

**On in vivo data**

We evaluated the RNN and MLP estimator output vs NLLS in voxels of the three white matter ROIs of rat cohort1 (*N*=56, pooled), which were never used for drawing training distributions. We express the relative difference between DL and NLLS estimations for each voxel as $diff = \frac{DL - NLLS}{NLLS} \times 100\,\%$ and $diff \leq 5\,\%$ is considered as DL and NLLS estimation agreement on that voxel. We further evaluated the agreement between these methods after additional filtering as explained above. Moreover, in order to test whether the RNN fitting would add additional correlation between parameters, we measured the Pearson correlation between *f* and the other four parameters in the three kinds of estimation on the rat dataset. Additionally, to test the generalization ability of the DL models trained on distributions drawn from rat in vivo data, we evaluated them on the HCP human datasets (*N*=6, pooled) for voxels with FA > 0.2.

Finally, WMTI parametric maps fitted by the proposed RNN approach were compared to the ones obtained from the conventional NLLS fitting approach for both rat and human data.

### 2.7. Analyzing rat group differences in WMTI parameters

In the end, differences in WMTI-Watson parameters in white matter between STZ and CTL groups in rat cohort1 were evaluated at each timepoint, based on the estimation of NLLS and RNN fitting approaches, respectively. Statistical test was performed as follows: 1) average WMTI model parameters in each white matter ROI were compared between STZ and CTL groups using two-tailed Mann-Whitney *U* test, at $\alpha$=0.05 significance; 2) longitudinal alterations within each group were calculated by one-way ANOVA and Tukey-Cramer correction, at $\alpha$= 0.05 significance (Tristão Pereira et al., 2021). Significant group differences based on the two estimations were compared.

### 3. Results

### 3.1 Estimation on synthetic data with similar distributions

Estimates from all three fitting approaches reached a good agreement with the ground-truth target on the synthetic dataset *Testset1*, which has the same distribution (Figure 3C) as the training dataset. Most datapoints lay along the identity line in the scatter plots (Figure 5A) and over 95% of the estimates had an error < 5% for each parameter (Figure 5B). However, noticeably, scatter plots of ground-truth vs. NLLS estimation display multiple "off-diagonal" clusters which indicate NLLS converged to inaccurate solutions especially when parameters hit their bounds (e.g. $D_a$, $D_{e,\parallel}$ and $c_2$), as a result of the existence of local minima in its fitting landscape. DL-based approaches, in contrast, were more likely to give accurate solutions with less extreme values or outliers even on boundaries (Figure 5A, RNN and MLP). Although NLLS outperforms RNN and MLP in terms of estimates with error < 1%, showing that NLLS tends to give better precision in its estimation, it also has more estimates with an error > 10% indicating reduced robustness in fitting than DL approaches (Figure 5B). Remarkably, if we consider an estimate with 5% of the GT to be accurate, RNN estimation delivers better accuracy than NLLS and MLP (Figure 5B).

More importantly, DL-based fitting approaches were approximately 5,000 times faster than NLLS estimation on the same computer. For example, NLLS took more than 10 hours to estimate 4 x $10^5$ data points using 32 CPU cores in parallel, while RNN- and MLP-based methods only took 8 and 5 seconds, respectively, on the GPU of the same machine (Table 1). The DL-based methods thus showed great advantage over NLLS in terms of computation time.

For completeness, the evaluation of DL models with five DKI metrics excluding MD as input can be found in *Supplementary Table 2*. Besides, the evaluation of DL models trained on datasets with only one group of distribution is provided in the *Supplementary Table 3* and *4*.

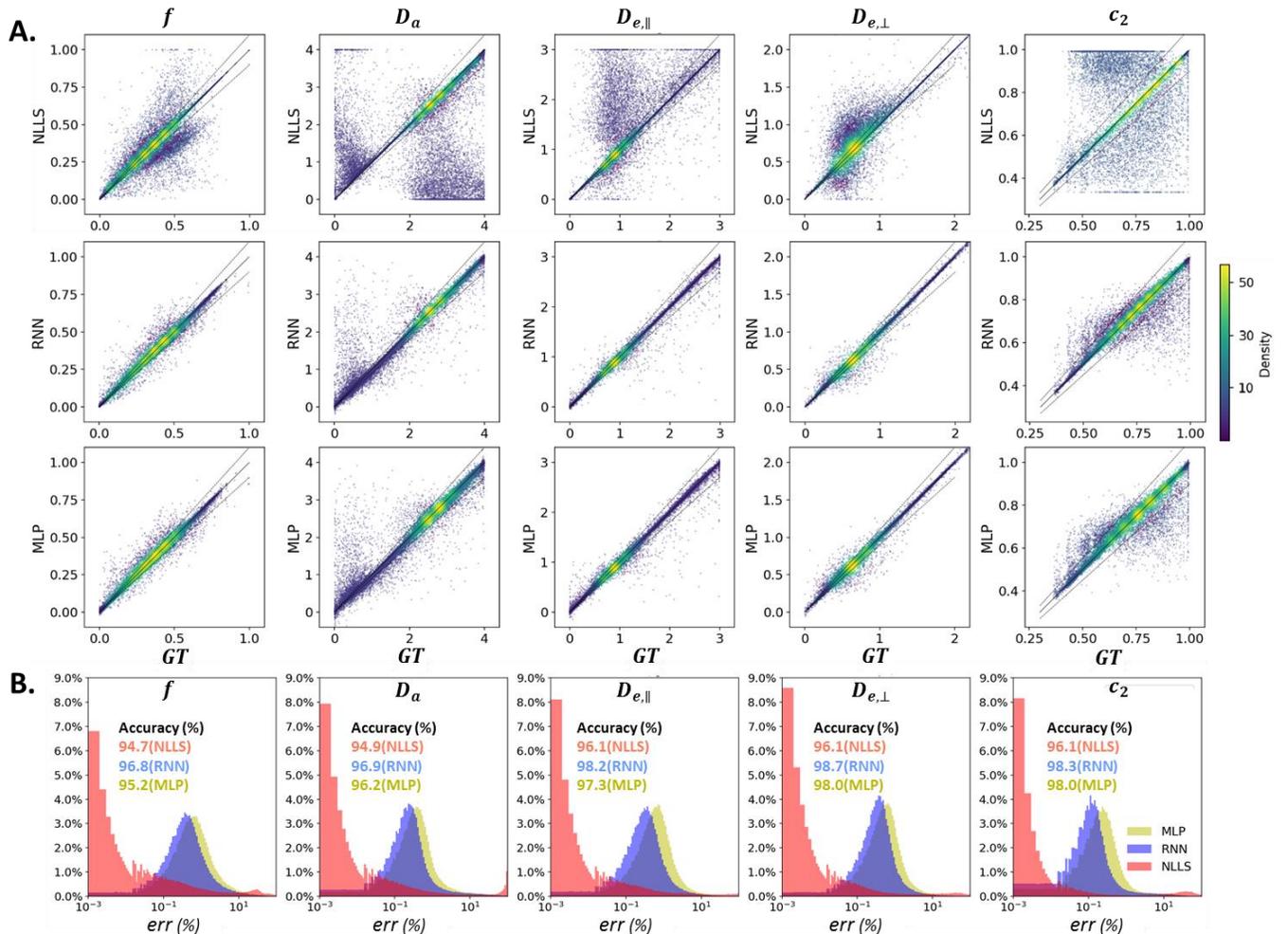

Figure 5. Performance of NLLS, RNN and MLP estimators on synthetic test dataset (*Testset*1). **A**: Scatter plots of the ground-truth targets vs. NLLS, RNN and MLP estimations for WMTI-Watson model parameters, overlaid by density maps. Black dashed lines indicate ±10% error. **B**: Error distributions of NLLS estimations (red), RNN (blue) and MLP predictions (yellow). The estimation accuracy (percentage of *err*<5%) of the three methods is quantified for each parameter. Although the majority of datapoints lay along the identity line, indicating a good agreement with the ground truth for all three fitting strategies, there were more off-diagonal outliers in the NLLS estimations than RNN and MLP, probably due to local minima in its fitting landscape and sensitivity to initialization. From the error histograms, NLLS works either extremely well, having the most estimations within 1% error, or extremely poorly, with more estimations having *err*>10% than RNN and MLP. When considering *err*<5%, the RNN model obtained the highest accuracy.

Table 1. Fitting time taken by the three methods on *Testset*1. NLLS fitting approach was implemented with Matlab's built-in function lsqnonlin with 32 workers in parallel while RNN and MLP networks were implemented with Pytorch and run on a NVIDIA graphic card on the same computer.

| Method | Data size | Fitting time [s] | Resource |
|---|---|---|---|
| NLLS | $4 \times 10^5$ | $3.83 \times 10^4$ | MATLAB, 32-core Intel Xeon 2.1 GHz CPU with 126 GB memory |
| RNN |  | 8 | Pytorch, 1 NVIDIA GeForce RTX 2080Ti graphic card with 11 GB memory |
| MLP |  | 5 |  |

## 3.2 Estimation on data with mismatched distributions

In contrast to *Testset1*, *Testset2* has more realistic distributions (Figure 4), drawn from NLLS estimates on in vivo rat data filtered only based on acceptable DKI values: this test dataset thus had more WMTI-Watson values hitting the bounds with NLLS estimation. As a result, the distribution matching technique was employed for RNN and MLP fitting (**Section 2.5**). However, results in Table 2 show that estimation accuracy for the three methods was lowered down by 5 to 10 % in *Testset*2 compared to *Testset*1. After further filtering for physically acceptable model parameter values either based on the estimates or on the ground truth, about 80% of data was retained and the estimation accuracy was significantly improved such that it was comparable to that in *Testset1*, especially for RNN fitting. In all cases, the RNN-based approach had the highest accuracy while the MLP network failed to outperform NLLS fitting for most parameters. Additionally, the evaluation of model performance without distribution matching can be found in *Supplementary Figure 1*.

Furthermore, the RNN-based and MLP-based approaches were compared to NLLS fitting in white matter voxels of the rat experimental dataset (cohort1). The RNN network achieved good agreement with NLLS (>92%) after filtering, which was substantially higher than MLP network (>55%) (Table 3). Furthermore, the parametric maps in white matter estimated from RNN and NLLS displayed good agreement between the two methods (Figure 6), which is consistent with Table 3. In addition, scatter plots and the correlation between $f$ and the other four parameters can be found in *Supplementary Figure 2*, which shows the RNN fitting doesn't increase the correlation among parameters after filtering out the unphysical values compared to NLLS and MLP.

Table 2. Accuracy performance of NLLS, RNN and MLP approaches. Performance was evaluated on the second synthetic test data (*Testset2*) which had more realistic distributions. Estimation accuracy was defined as the percentage of data with an error < 5% between estimations and the ground-truth target. Performance was also evaluated on VOIs obtained by a filter (0.01 < f < 1; 0.01< $D_a$ < 3.9; 0.01 < [ $D_{e,\|}, D_{e,\perp}$] < 3; 0.33 < $c_2$ < 1; $D_a > D_{e,\|}$) based on the three estimations and ground truth, respectively. The relative data size of VOIs is shown. Scatter plots of estimations vs the ground truth can be found in *Supplementary Figure 3*. (Embedding ratio: γ = 10, test batch size = 2048)

| Method | Estimation Accuracy (%) | | | | | Relative size (%) | Filtering |
|---|---|---|---|---|---|---|---|
| | $f$ | $D_a$ | $D_{e,\|}$ | $D_{e,\perp}$ | $c_2$ | | |
| NLLS | 85.5 | 86.7 | 89.5 | 90.4 | 92.1 | 100 | NO |
| **RNN** | **87.9** | **89.9** | **91.5** | **92.9** | **95.4** | **100** | |
| MLP | 84.3 | 87.7 | 87.3 | 89.7 | 94.5 | 100 | |
| NLLS | 89.9 | 92.3 | 93.8 | 94.6 | 96.3 | 81.3 | Based on own estimations |
| **RNN** | **91.3** | **94.7** | **93.9** | **95.0** | **98.1** | **82.0** | |
| MLP | 86.4 | 92.2 | 89.7 | 91.5 | 97.6 | 83.8 | |
| NLLS | 92.3 | 94.4 | 94.8 | 95.4 | 97.0 | 79.0 | Based on ground truth |
| **RNN** | **93.7** | **97.0** | **95.9** | **96.7** | **98.9** | **79.0** | |
| MLP | 89.5 | 95.3 | 92.4 | 93.5 | 98.4 | 79.0 | |

Table 3. Agreement between NLLS and the two DL fitting approaches on rat data. Comparison was performed in the three white matter ROIs in rat cohort1 datasets (*N*=56). The estimation difference for each voxel was defined as $diff = \frac{DL - NLLS}{NLLS} \times 100\ \%$ and $diff \leq 5\ \%$ means DL and NLLS agree on that voxel. This table shows the percentage of voxels having an estimation agreement with NLLS for RNN and MLP. The agreement was also evaluated in VOIs by filtering (0.01 < f < 1; 0.01< $D_a$ < 3.9; 0.01 < [ $D_{e,\parallel}$, $D_{e,\perp}$] < 3; 0.33 < $c_2$ < 1; $D_a > D_{e,\parallel}$) based on either NLLS or DL outcomes, and the relative size of retained voxels after filtering is shown. (Embedding ratio: γ = 10, test batch size = 2048)

| Method | Agreement with NLLS (%) | | | | | Relative size (%) | Filtering |
|---|---|---|---|---|---|---|---|
| | $f$ | $D_a$ | $D_{e,\parallel}$ | $D_{e,\perp}$ | $c_2$ | | |
| **RNN** | **72.5** | **75.3** | **79.9** | **78.5** | **80.4** | 100 | NO |
| MLP | 46.6 | 55.8 | 54.6 | 59.8 | 79.4 | 100 | |
| **RNN** | **92.6** | **94.2** | **95.2** | **95.1** | **95.9** | 75.9 | Based on DL estimations |
| MLP | 55.3 | 63.6 | 62.6 | 69.7 | 90.1 | 83.4 | |
| **RNN** | **95.9** | **97.6** | **98.1** | **98.2** | **99.8** | 72.6 | Based on NLLS estimations |
| MLP | 61.7 | 71.3 | 68.7 | 77.0 | 97.4 | 72.6 | |

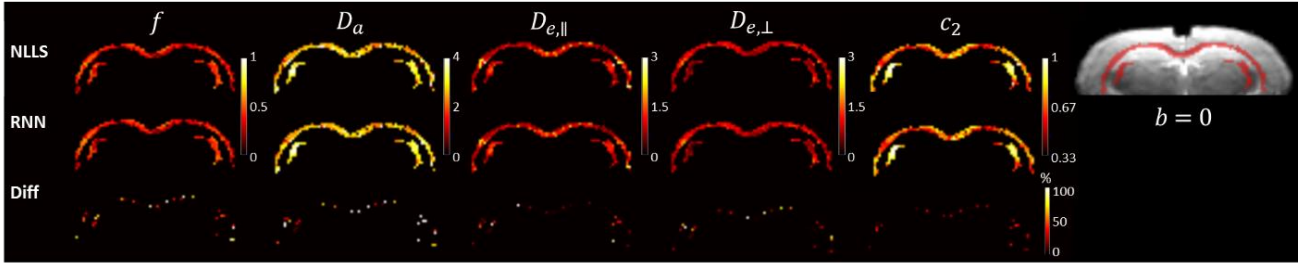

Figure 6. WMTI-Watson parameter maps of white matter regions in one subject from rat cohort1. Maps in the first and second row were reconstructed based on NLLS and RNN estimations, respectively. The third row was the relative difference in percentage between NLLS and RNN. The last column shows an image of the same slice with b-value = 0 overlaid by a mask of WM regions in red, for anatomical reference.

### 3.3 Generalization on human data

Trained on synthetic data with distributions drawn from the rat brain, the generalization ability of the RNN and MLP models was tested on a pooled HCP dataset including six subjects. The estimation agreement of DL models with NLLS in human data (Table 4) is even higher than in rat data (Table 3). One reason might be the high image quality in the HCP dataset. Furthermore, the RNN model outperformed the MLP model significantly. Excellent RNN agreement with NLLS before (>80.9%) and after filtering (94.9%) indicates that the RNN encoder-decoder model has a good generalization ability on various data, without retraining the model. Moreover, the parametric maps in regions with FA>0.2 were estimated and compared for RNN and NLLS (Figure 7). The largest overall discrepancy between RNN and NLLS estimation was found for $D_a$.

Table 4. Same as Table 3. Agreement between NLLS and two DL fitting approaches on the HCP data. Evaluation was performed in voxels with FA>0.2 and datasets (*N*=6) were pooled. (Embedding ratio: γ = 10, test batch size = 2048)

| Method | Agreement with NLLS (%) | | | | | Relative size (%) | Filtering |
|---|---|---|---|---|---|---|---|
| | $f$ | $D_a$ | $D_{e,\parallel}$ | $D_{e,\perp}$ | $c_2$ | | |
| **RNN** | **80.9** | **82.5** | **86.3** | **84.2** | **85.8** | **100** | NO |
| MLP | 66.2 | 69.8 | 56.5 | 62.5 | 85.8 | 100 | |
| **RNN** | **94.9** | **96.2** | **96.3** | **96.3** | **97.3** | **83.7** | Based on DL estimations |
| MLP | 74.5 | 77.9 | 61.5 | 68.8 | 94.3 | 88.2 | |
| **RNN** | **97.8** | **99.2** | **99.0** | **99.1** | **99.8** | **80.8** | Based on NLLS estimations |
| MLP | 80.1 | 84.1 | 65.6 | 73.1 | 99.3 | 80.8 | |

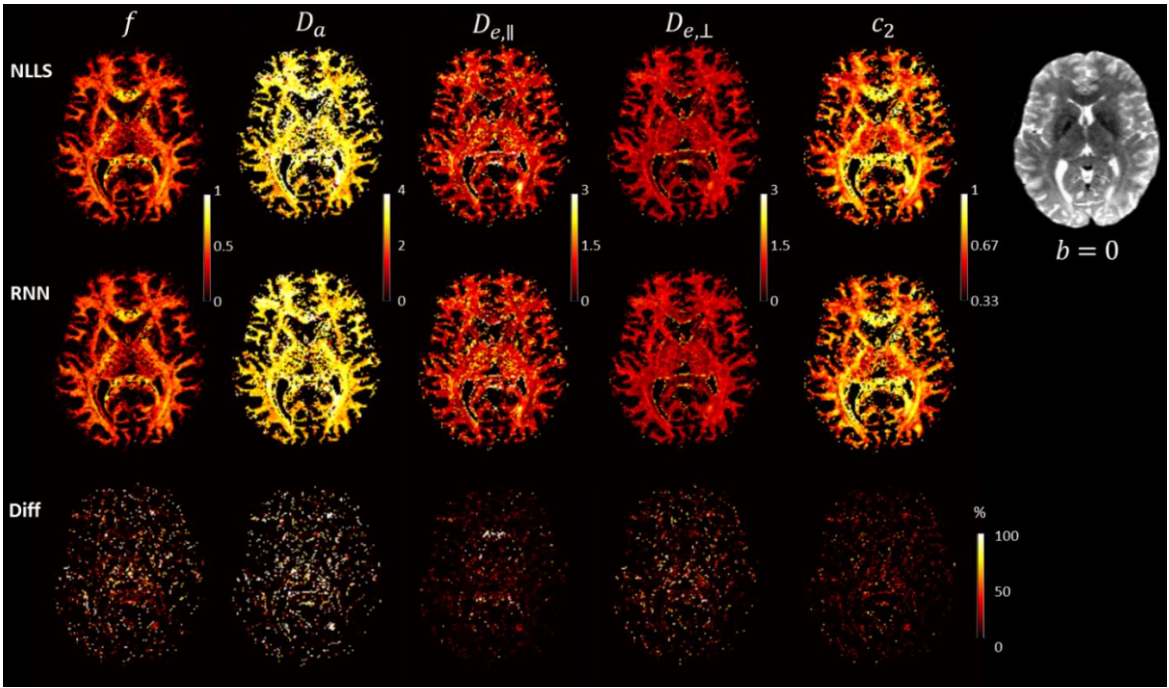

Figure 7. WMTI parameter maps of regions with FA>0.2 in one subject from the HCP dataset. Maps in the first and second row were reconstructed based on NLLS and RNN estimations, respectively. The third row was the relative difference in percentage between NLLS and RNN. The last column shows an image of the same slice with b-value = 0.

### 3.4 Group difference detection in rat data

Figure 8 shows group difference in terms of WMTI model parameters in corpus callosum between STZ and CTL rats in cohort1, detected using NLLS (Figure 8A) and RNN (Figure 8B) estimates, respectively. The overall results are remarkably similar, with reduced intra-axonal diffusivity $D_a$ (at timepoints 2 and 13 weeks) and axonal fraction $f$ (at 13 weeks) in the STZ group. Increased extra-axonal diffusivity perpendicularly to axons ($D_{e,\perp}$) in STZ rats was observed in both cases but the difference was significant only using RNN estimates. Notably, $D_a$ estimates were higher using NLLS, similar to the trend in HCP data. Results of group difference in CG and Fi can be found in *Supplementary Figure 4*.

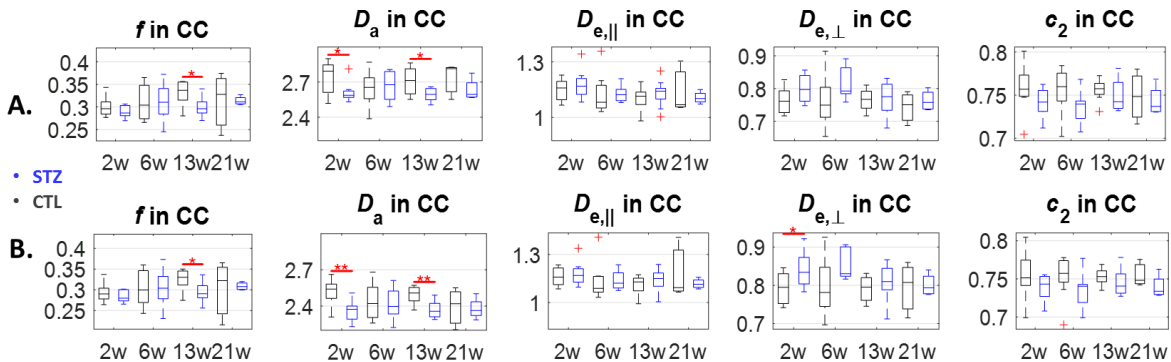

Figure 8. Differences in WMTI parameters in corpus callosum (CC) between STZ and CTL groups at 2, 6, 13 and 21 weeks in rat cohort1. Group differences were analyzed based on the WMTI model fitted by NLLS (**A**) and RNN (**B**), respectively. The number of datasets in CTL/STZ group for each timepoint was: 8/8, 8/8, 8/9 and 3/4. Two-tailed Mann-Whitney U test for inter-group comparison (red bars) and one-way ANOVA with Tukey-Cramer correction for within-group comparison across time. ∗ : p < 0.05, ∗ ∗ : p < 0.01. + : outlier values (but not excluded from the analysis).

## 4. Discussion and conclusion

Conventional model fitting strategies (e.g. NLLS) are confronted with many problems such as solution degeneracy due to noise and local minima in the shallow fitting landscape (Jelescu et al., 2020; Novikov et al., 2019) and extremely slow estimation speed. DL-based fitting approaches showed promising outcomes in tackling those problems. For example, the fitting speed was improved dramatically, reducing the computation time from hours to seconds (Table 1) and thus enabling model estimation in real time or for large-scale datasets. Furthermore, scatter plots of the GT target vs estimations in Figure 5 demonstrated that DL-based fitting produced more robust estimation than NLLS in the degenerate cases. These findings are consistent with reports from other dMRI modelling studies using deep learning for parameter estimation (Barbieri et al., 2020; Bertleff et al., 2017; de Almeida Martins et al., 2021; Jelescu et al., 2022; Park et al., 2021; Ye, 2017).

However, as end-to-end (E2E) learning models, the DL-based solvers in previous studies directly take the dMRI signal as input to estimate microstructure parameters. The advantage of this approach is its straightforwardness and simplicity in real-world application. However, E2E models are more likely to have bad generalizability to dMRI signals with different noise levels and artefacts, acquired using different protocols and from different species (e.g. human, rat or mouse). For example, a CNN-based solver (Park et al., 2021) that is trained on human brain data is less like to generalize to rat brain due to different brain structures since a CNN model relies strongly on structural information in its prediction. As a consequence, training of the DL model has to be repeated for different acquisition protocols or anatomical regions (Barbieri et al., 2020).

Instead of fitting a WM model directly to the dMRI signal, WMTI-Watson has been proposed as an alternative to extract model parameters from DKI tensor metrics (Fieremans et al., 2011; Jensen et al., 2017; Jespersen et al., 2018), which are derived linearly from the dMRI signal. While this framework has known limitations, more notably the limited sensitivity due to the low b-value range, the axially-symmetric ODF assumption and the use of linear diffusion encoding only (Afzali et al., 2019; Coelho et al., 2022, 2019; Lampinen et al., 2020), it also presents several advantages. Firstly, the acquisition protocol needed to estimate DKI is clinically feasible and increasingly widespread in clinical protocols – combined DTI and DKI estimation are becoming part of standard processing pipelines for dMRI data. Second, from the perspective of DL fitting, the variability introduced by noise level, acquisition protocol or anatomical region is contained within the DKI estimation – provided the data distributions can be matched between training and test set, the model can therefore be applied directly without retraining.

As a result, to capitalize on the generalization potential of the trained DL model for WMTI-Watson estimation, we employed a distribution matching technique when applying the trained model to new test data with a potentially different distribution from the training data. The matching was achieved by randomly embedding the new data in the training data with an embedding ratio (e.g. 1:10) before inference, to ensure the data fed to the DL model had a similar distribution to the training data. Finally, the estimations for the new test data were retrieved from the recorded positions in the embedded data.

Notably, the DL solver was trained on synthetic data with GT model parameters whose distribution was drawn from in vivo rat data with NLLS fit. It is worth pointing out that WMTI-Watson can have two mathematical solutions $D_a < D_{e,\parallel}$ or $D_a > D_{e,\parallel}$ whereas we chose the second one. When designing the training dataset, we chose to mix the two types of data (I. with both solutions; II. with only $D_a > D_{e,\parallel}$) with some ratio (e.g. I/II = 1/3) to achieve better performance than comprising only one data type. Indeed, although models trained exclusively on data II might have the best accuracy in the expected WM features, their solution space is restricted which may reduce their performance in non-typical tissue structures, including pathology. Incorporating a small fraction of data I type can increase the fitting landscape and the solution space, thus improving model generalizability. This finding was also reported in another study (de Almeida Martins et al., 2021).

With the optimized training dataset and distribution matching, the proposed RNN encoder-decoder network achieved better accuracy than NLLS and MLP on the synthetic dataset with realistic distributions (Table 2). Remarkably, trained on the same dataset, the RNN-based solver substantially outperformed the MLP-based solver in terms of agreement with NLLS estimation on the in vivo rat data (Table 3) and human data (Table 4), which displayed the better performance and generalizability of the RNN network being a more advanced architecture. Based on LSTM and employing the attention mechanism, the RNN encoder-decoder has demonstrated great success in machine translation and seq2seq prediction (Cho et al., 2014; Sutskever et al., 2014) with various sentence or sequence length. We believe this RNN model can be applied to other model estimation problems apart from the WMTI-Watson model, where a sequence of parameters is to be predicted from another sequence even though the parameters are independent from each other. Indeed, no spurious correlation was found in the RNN estimation after filtering out the unphysical values.

We note that the largest discrepancy between NLLS and RNN fitting was found in $D_a$. This could be due to the fact that $D_a$ is the most difficult parameter to estimate in such models (Jelescu et al., 2016; Novikov et al., 2018; Palombo et al., 2020) and can be highly affected by noise. In NLLS, its estimate is more likely to hit the upper bound if the fit does not converge properly, while the RNN estimate tends to remain close to the prior mean.

Finally, we highlight that, thanks to distribution-matching through embedding, the proposed RNN-based fitting approach has a good transferability to data of different species or anatomical regions for WMTI-Watson model estimation. Trained on synthetic data with distribution extracted from in vivo rat data, it obtained high estimation agreement with NLLS and valid parameter maps not only on independent rat data (Figure 6) but also on human data (Figure 7) with much improved speed. The RNN solver was further validated by the group comparison between STZ and CTL rats in the rat cohort1. The latter showed similar group differences were found when using either the NLLS or RNN estimation (Figure 8), suggesting RNN preserves comparable sensitivity to parameter changes with pathology as NLLS.

For other biophysical models where the dMRI signal is fit directly, the direct reuse of a given network without retraining is limited due to strong dependence not just on the parameter distributions – which can be mitigated by embedding – but also on the acquisition protocol and noise level. Transfer learning algorithms may be better suited in such cases (Banerjee et al., 2018; Chen et al., 2020; Valverde et al., 2021). We also note that our current RNN does not learn or model noise explicitly. Noise propagation from the dMRI measurements to the DTI and DKI scalar metrics is not straightforward and depends also on the number of measurements available for the tensor estimation (hence on the acquisition protocol). It is assumed that the impact of noise is mitigated in the weighted linear least square estimation for DKI (Veraart et al., 2013).

In conclusion, we proposed a novel DL solver based on the RNN encoder-decoder network to replace the conventional NLLS fitting for WMTI-Watson model estimation. On the basis of LSTM and using the attention mechanism, this RNN network is able to predict long sequences in a seq2seq problem. Starting from the DKI tensor and employing distribution matching, the RNN estimator can be easily applied to various datasets be they animal or human data without re-training, which shows great potential for a widespread use. Our study showed the RNN-based fitting approach achieved high estimation accuracy in the synthetic data and good agreement with NLLS on in vivo data especially in VOIs including only voxels within valid parameter intervals and the desired solution. Foremost, with thousands of times less in fitting time, we believe the proposed RNN encoder-decoder network can readily replace NLLS in parameter estimation of WMTI-Watson model.

**Code availability**

The code is available on https://github.com/Mic-map/WMTI-Watson_DL.

**Acknowledgements**

The authors thank Catarina Tristão Pereira for contributing code for WM ROI segmentation and acknowledge access to the facilities and expertise of the CIBM Center for Biomedical Imaging, a Swiss research center of excellence founded and supported by Lausanne University Hospital (CHUV), University of Lausanne (UNIL), Ecole Polytechnique Fédérale de Lausanne (EPFL), University of Geneva (UNIGE) and Geneva University Hospitals (HUG). I.O.J. is supported by the Swiss National Science Foundation Eccellenza Fellowship PCEFP2_194260.

**Supplementary Material** for "Parameter estimation for WMTI-Watson model of white matter using encoder-decoder recurrent neural network"

*Supplementary Table 1.* Details of NLLS fitting using Matlab's built-in function lsqnonlin. $\kappa$: concentration parameter for the Watson distribution, directly related to $c_2$.

|  | Bound | $f$ | $D_a$ | $D_{e,\parallel}$ | $D_{e,\perp}$ | $\kappa$ |
|---|---|---|---|---|---|---|
| Box constraints | Lower | 0 | 0 | 0 | 0 | 0 |
|  | Upper | 1 | 4 | 3 | 3 | 128 |
| Random Initialization | Lower | 0.1 | 1.5 | 0.5 | 0 | 4 |
|  | Upper | 0.9 | 2.5 | 1.5 | 0.5 | 16 |

*Supplementary Table 2.* Accuracy performance of NLLS, RNN and MLP approaches. RNN and MLP models were trained on the dataset using only five DKI metrics as input with MD excluded. Performance was evaluated on the second synthetic test data (*Testset2*) which had more realistic distributions. Estimation accuracy was defined as the percentage of data with an error < 5% between estimations and the ground-truth target. Performance was also evaluated on VOIs obtained by a filter (0.01 < f < 1; 0.01< $D_a$ < 3.9; 0.01 < [ $D_{e,\parallel}, D_{e,\perp}$ ] < 3; 0.33 < $c_2$ < 1; $D_a > D_{e,\parallel}$) based on the three estimations and ground truth, respectively. The relative data size of VOIs is shown. Embedding ratio: $\gamma$ = 10, test batch size = 2048)

| Method | Estimation Accuracy (%) | | | | | Relative size (%) | Filtering |
|---|---|---|---|---|---|---|---|
|  | $f$ | $D_a$ | $D_{e,\parallel}$ | $D_{e,\perp}$ | $c_2$ |  |  |
| NLLS | 85.5 | 86.7 | 89.5 | 90.4 | 92.1 | 100 | NO |
| **RNN** | **86.9** | **89.4** | **90.4** | **91.9** | **95.3** | **100** |  |
| MLP | 84.3 | 87.9 | 87.3 | 89.5 | 94.5 | 100 |  |
| NLLS | 89.9 | 92.3 | 93.8 | 94.6 | 96.3 | 81.3 | Based on own estimations |
| **RNN** | **89.8** | **94.0** | **92.6** | **93.8** | **98.0** | **83.3** |  |
| MLP | 86.6 | 92.3 | 89.6 | 91.2 | 97.5 | 83.9 |  |
| NLLS | 92.3 | 94.4 | 94.8 | 95.4 | 97.0 | 79.0 | Based on ground truth |
| **RNN** | **92.5** | **96.5** | **95.3** | **95.7** | **98.8** | **79.0** |  |
| MLP | 89.7 | 95.4 | 92.4 | 93.3 | 98.4 | 79.0 |  |

*Supplementary Table 3.* Accuracy performance of NLLS, RNN and MLP approaches. RNN and MLP models were trained on the dataset following only the **first** group of distribution. Performance was evaluated on the second synthetic test data (*Testset2*) which had more realistic distributions. Estimation accuracy was defined as the percentage of data with an error < 5% between estimations and the ground-truth target. Performance was also evaluated on VOIs obtained by a filter $(0.01 < f < 1; 0.01 < D_a < 3.9; 0.01 < [D_{e,\parallel}, D_{e,\perp}] < 3; 0.33 < c_2 < 1; D_a > D_{e,\parallel})$ based on the three estimations and ground truth, respectively. The relative data size of VOIs is shown (Embedding ratio: $\gamma = 10$, test batch size = 2048)

| Method | Estimation Accuracy (%) | | | | | Relative size (%) | Filtering |
|---|---|---|---|---|---|---|---|
| | $f$ | $D_a$ | $D_{e,\parallel}$ | $D_{e,\perp}$ | $c_2$ | | |
| NLLS | 85.5 | 86.7 | 89.5 | 90.4 | 92.1 | 100 | None |
| **RNN** | **82.6** | **86.0** | **89.3** | **90.2** | **94.3** | **100** | |
| MLP | 83.6 | 83.9 | 87.1 | 89.0 | 93.2 | 100 | |
| NLLS | 89.9 | 92.3 | 93.8 | 94.6 | 96.3 | 81.3 | Based on own estimations |
| **RNN** | **85.7** | **91.2** | **91.3** | **92.6** | **97.4** | **82.6** | |
| MLP | 86.8 | 89.6 | 89.8 | 91.9 | 96.9 | 82.6 | |
| NLLS | 92.3 | 94.4 | 94.8 | 95.4 | 97.0 | 79.0 | Based on ground truth |
| **RNN** | **87.6** | **92.4** | **95.9** | **93.7** | **97.1** | **79.0** | |
| MLP | 88.4 | 90.5 | 91.5 | 92.5 | 96.5 | 79.0 | |

*Supplementary Table 4.* Accuracy performance of NLLS, RNN and MLP approaches. RNN and MLP models were trained on the dataset following only the **second** group of distribution. Performance was evaluated on the second synthetic test data (*Testset2*) which had more realistic distributions. Estimation accuracy was defined as the percentage of data with an error < 5% between estimations and the ground-truth target. Performance was also evaluated on VOIs obtained by a filter $(0.01 < f < 1; 0.01 < D_a < 3.9; 0.01 < [D_{e,\parallel}, D_{e,\perp}] < 3; 0.33 < c_2 < 1; D_a > D_{e,\parallel})$ based on the three estimations and ground truth, respectively. The relative data size of VOIs is shown (Embedding ratio: $\gamma = 10$, test batch size = 2048)

| Method | Estimation Accuracy (%) | | | | | Relative size (%) | Filtering |
|---|---|---|---|---|---|---|---|
| | $f$ | $D_a$ | $D_{e,\parallel}$ | $D_{e,\perp}$ | $c_2$ | | |
| NLLS | 85.5 | 86.7 | 89.5 | 90.4 | 92.1 | 100 | None |
| **RNN** | **85.7** | **90.3** | **89.2** | **91.2** | **95.7** | **100** | |
| MLP | 84.3 | 89.0 | 86.9 | 90.4 | 95.0 | 100 | |
| NLLS | 89.9 | 92.3 | 93.8 | 94.6 | 96.3 | 81.3 | Based on own estimations |
| **RNN** | **88.2** | **94.0** | **91.3** | **92.7** | **98.1** | **84.6** | |
| MLP | 86.0 | 92.3 | 89.2 | 91.7 | 97.5 | 85.3 | |
| NLLS | 92.3 | 94.4 | 94.8 | 95.4 | 97.0 | 79.0 | Based on ground truth |
| **RNN** | **92.3** | **97.9** | **95.0** | **95.7** | **99.6** | **79.0** | |
| MLP | 90.2 | 97.0 | 92.9 | 94.7 | 99.4 | 79.0 | |

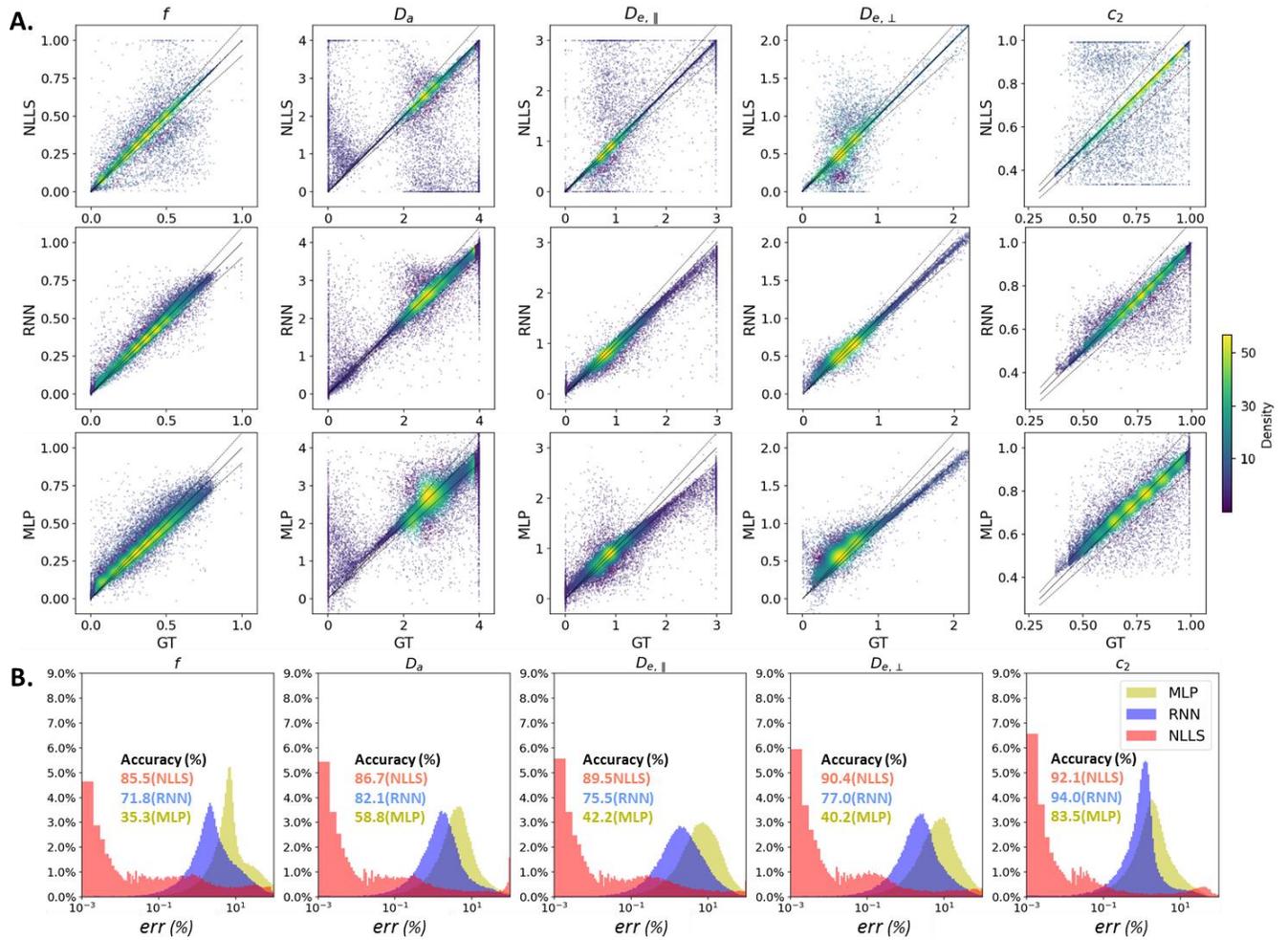

*Supplementary Figure 9.* Performance of NLLS, RNN and MLP estimators on synthetic test dataset (*Testset2*). **Distribution matching was not used for RNN and MLP here**. No filtering was applied in the estimation. **A**: Scatter plots of the ground-truth targets vs. NLLS, RNN and MLP estimations for WMTI-Watson model parameters, overlaid by density maps. Black dashed lines indicate ±10% error. **B**: Error distributions of NLLS estimations (red), RNN (blue) and MLP predictions (yellow). The estimation accuracy (percentage of *err*<5%) of the three methods is quantified for each parameter.

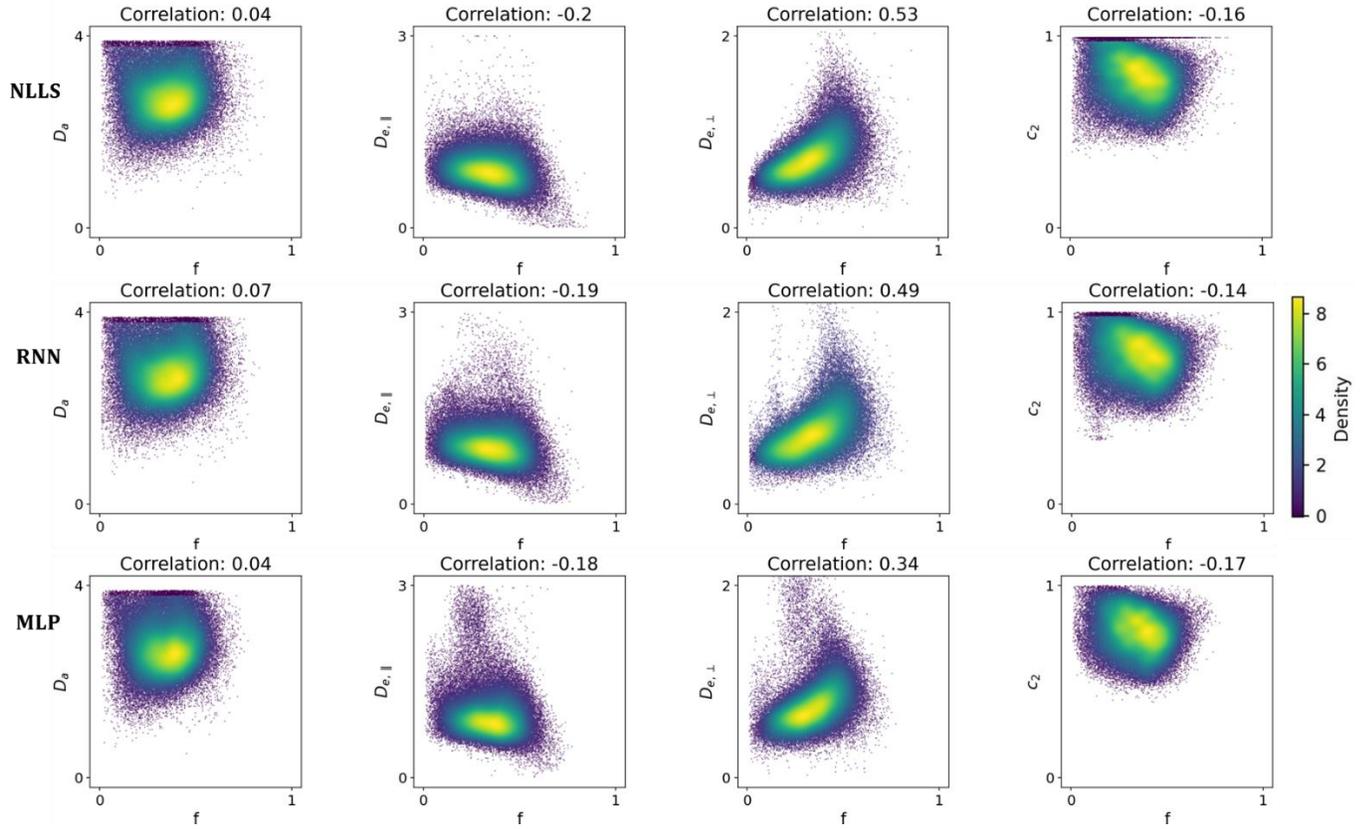

*Supplementary Figure 2. Scatter plots and the Pearson correlation coefficient of $f$ vs $D_a$, $D_{e,\parallel}$, $D_{e,\perp}$ and $c_2$ in NLLS, RNN and MLP estimation of rat WM in vivo data. Scatter plots were overlaid by density maps and unphysical values in the estimation have been filtered out according to the following constraints: $0.01 < f < 1$; $0.01 < D_a < 3.9$; $0.01 < [D_{e,\parallel}, D_{e,\perp}] < 3$; $0.33 < c_2 < 1$; $D_a > D_{e,\parallel}$.*

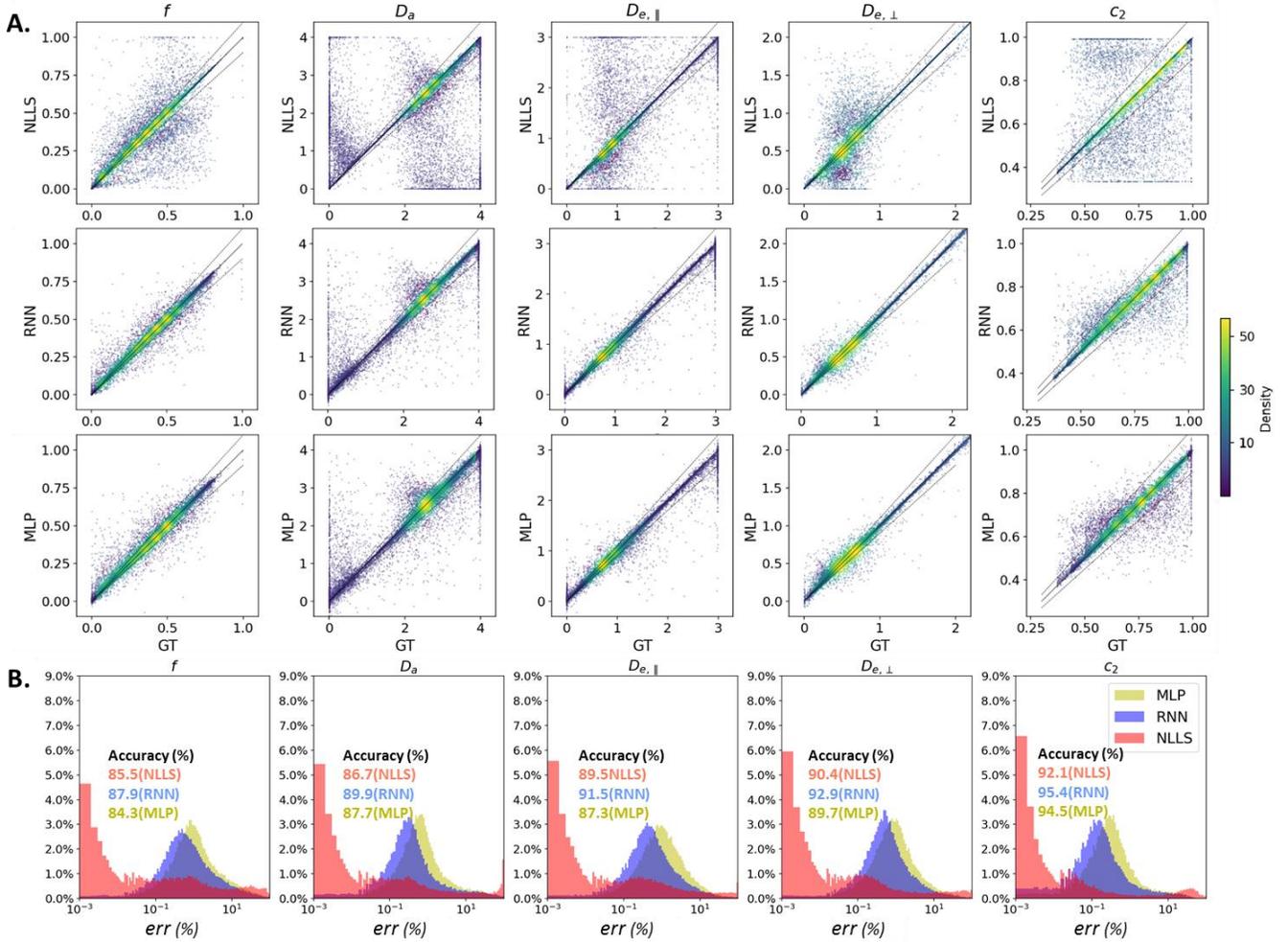

*Supplementary Figure 3.* Performance of NLLS, RNN and MLP estimators on synthetic test dataset (*Testset2*). **Distribution matching was used for RNN and MLP**. No filtering was applied in the estimation. **A**: Scatter plots of the ground-truth targets vs. NLLS, RNN and MLP estimations for WMTI-Watson model parameters, overlaid by density maps. Black dashed lines indicate ±10% error. **B**: Error distributions of NLLS estimations (red), RNN (blue) and MLP predictions (yellow). The estimation accuracy (percentage of *err*<5%) of the three methods is quantified for each parameter.

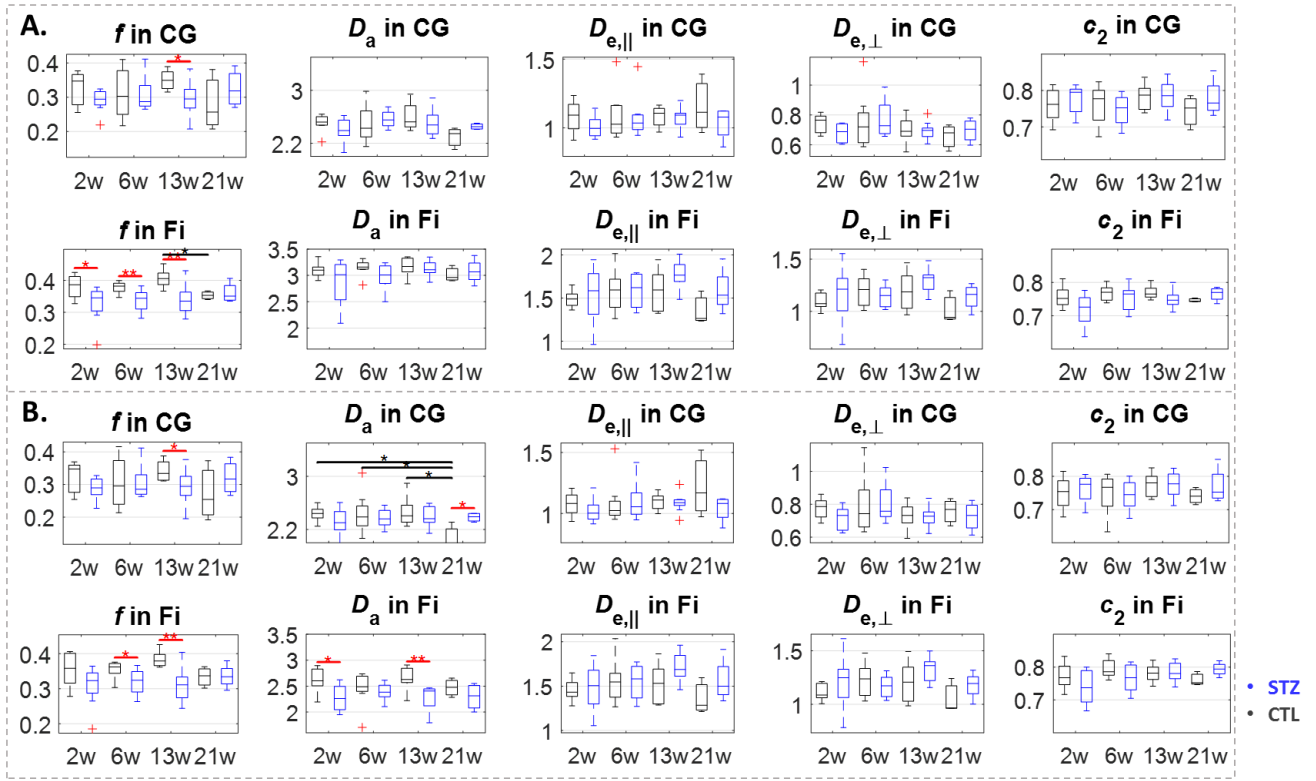

*Supplementary Figure 10.* Differences in WMTI parameters in Cingulum (CG) and Fimbria (Fi) between STZ and CTL groups at 2, 6, 13 and 21 weeks in rat cohort1. Group differences were analyzed based on the WMTI model fitted by NLLS (**A**) and RNN (**B**), respectively. The number of datasets in CTL/STZ group for each timepoint was: 8/8, 8/8, 8/9 and 3/4. Two-tailed Mann-Whitney U test for inter-group comparison (red bars) and one-way ANOVA with Tukey-Cramer correction for within-group comparison across time. ∗ : $p < 0.05$, ∗∗ : $p < 0.01$. + : outlier values (but not excluded from the analysis).